\documentstyle[12pt,aasms4]{article}
\begin{document}
\title{Evolution of Iron K$_{\alpha}$ Line Emission in the Black Hole
Candidate GX~339$-$4 }
\author{Y.X. Feng\altaffilmark{1}, S.N. Zhang\altaffilmark{2,3},
X. Sun\altaffilmark{2}, Ph. Durouchoux \altaffilmark{4}, 
Wan Chen\altaffilmark{5,6}, and Wei Cui\altaffilmark{1}}

\altaffiltext{1}{Department of Physics, Purdue University, West Lafayette, 
IN 47907; fengyx, cui@physics.purdue.edu}
\altaffiltext{2}{Physics Department, University of Alabama in 
Huntsville, Huntsville, AL 35899; zhangsn, sunxx@jet.uah.edu}
\altaffiltext{3}{Space Sciences Lab., SD50, NASA Mashall Space Flight Center, 
Huntsville, AL 35812}
\altaffiltext{4}{ CE-Saclay, DSM, DAPNIA, Service d'Astrophysique, Gif-sur
Yvette Cedex, France; \\  durvla@discovery.saclay.cea.fr}
\altaffiltext{5}{NASA/Goddard Space Flight Center, Code 661, Greenbelt, 
MD 20771; chen@milkyway.gsfc.nasa.gov}
\altaffiltext{6}{Dept. of Astronomy, University of Maryland, College Park, 
MD 20742}

\begin{abstract} 
GX 339-4 was regularly monitored with RXTE during a period (in 1999) 
when its X-ray flux decreased significantly (from 4.2$\times 10^{-10}$ 
erg cm$^{-2} s^{-1}$ to 7.6$\times 10^{-12}$ erg cm$^{-2}$s$^{-1}$ in 
the 3--20 keV band), as the source settled into the ``off state''. Our 
spectral analysis revealed the presence of a prominent iron K$_{\alpha}$ 
line in the observed spectrum of the source for all observations. The 
line shows an interesting evolution: it is centered at $\sim$6.4 keV when 
the measured flux is above 5$\times 10^{-11}$ erg cm$^{-2} s^{-1}$, but 
is shifted to $\sim$6.7 keV at lower fluxes. The equivalent width of the 
line appears to increase significantly toward lower fluxes, although it 
is likely to be sensitive to calibration uncertainties. While the 
fluorescent emission of neutral or mildly ionized iron atoms in the 
accretion disk can perhaps account for the 6.4 keV line, as is often 
invoked for black hole candidates, it seems difficult to understand the 
6.7 keV line with this mechanism, because the disk should be less ionized 
at lower fluxes (unless its density changes drastically). On the other 
hand, the 6.7 keV line might be due to recombination cascade of hydrogen 
or helium like iron ions in an optically thin, highly ionized plasma. 
We discuss the results in the context of proposed accretion models.
\end{abstract}


\section{Introduction}

GX~339$-$4 was discovered in the MIT OSO-7 sky survey (Markert et al. 
1973). We now know that for all practical purposes GX 339-4 is a 
persistent X-ray source. It is considered as a black hole candidate 
(BHC), because its spectral and temporal X-ray properties are similar 
to those of dynamically determined BHCs, such as Cyg X-1 (see review 
by Tanaka \& Lewin 1995). For instance, GX~339$-$4 shows a variety of 
spectral states, often designated as ``off'', ``low'', ``intermediate'', 
``high'', and ``very high''. It is worth noting, however, that the 
``off state'' is thought to be simply a weak ``low state'' (e.g., 
Motch et al. 1985; Kong et al. 2000). Recently, it is shown that 
GX 339-4 also bears resemblance to GRO~J1655$-$40 and GRS~1915$+$105, 
two well-known ``micro-quasars'', in terms of the observed radio 
properties (Zhang et al. 1997). The radio emission of GX 339-4 is also 
thought to be due to synchrotron emission of electrons 
in jet-like outflows (Fender et al. 1999). Interestingly, the radio 
emission, as well as the optical emission, is found to be strongly 
correlated with the hard X-ray emission but anti-correlated with
the soft X-ray emission (Fender et al. 1997, 1999; Corbel et al. 2000; 
Steiman-Cameron et al. 1990; Imamura et al. 1990; Ilovaisky et al. 
1986); similar correlations were first noted for Cyg X-1 (Zhang et 
al. 1997). 

The observed X-ray spectrum of GX 339-4 can often be described by a 
two-component model: a blackbody-like component at low energies and a 
power law at high energies, which is typical of BHCs (Tanaka \& Lewin 
1995). The soft component is usually attributed to emission from the 
hot, inner region of an optically thick and geometrically thin disk, 
and the hard component to the Comptonization of soft photons by 
energetic electrons (thermally or non-thermally distributed) in the 
region. Iron K$_{\alpha}$ line emission has been detected in GX 339-4
at around 6.4 keV (Ueda et al. 1994; Wilms et al. 1999; Smith et al. 
1999). The line probably originates in the fluorescent emission of 
cold iron atoms in the accretion disk, which is illuminated by the 
hard X-ray source. It can, therefore, be used as a valuable tool to 
study the physical conditions of accretion flows, such as their 
ionization states. 

In this paper, we report results from a detailed study of iron 
K$_{\alpha}$ line emission in GX 339-4. The source was observed by the 
instruments aboard the {\it Rossi X-ray Timing Explorer} (RXTE) over a 
period when its flux dropped by almost two orders of magnitude (down 
to the detection limit of RXTE). We show that not only was the line
always present in the observed X-ray spectrum but it evolved 
significantly at low fluxes (i.e., in the ``low'' or ``off'' states). 
We discuss the significance of the line detection, in light of possible 
uncertainties in the calibration of the instruments. We also discuss 
the potential impact of the results on the proposed accretion models for 
BHCs.

\section{Data and Data Reduction}
\begin{figure}
\plotone{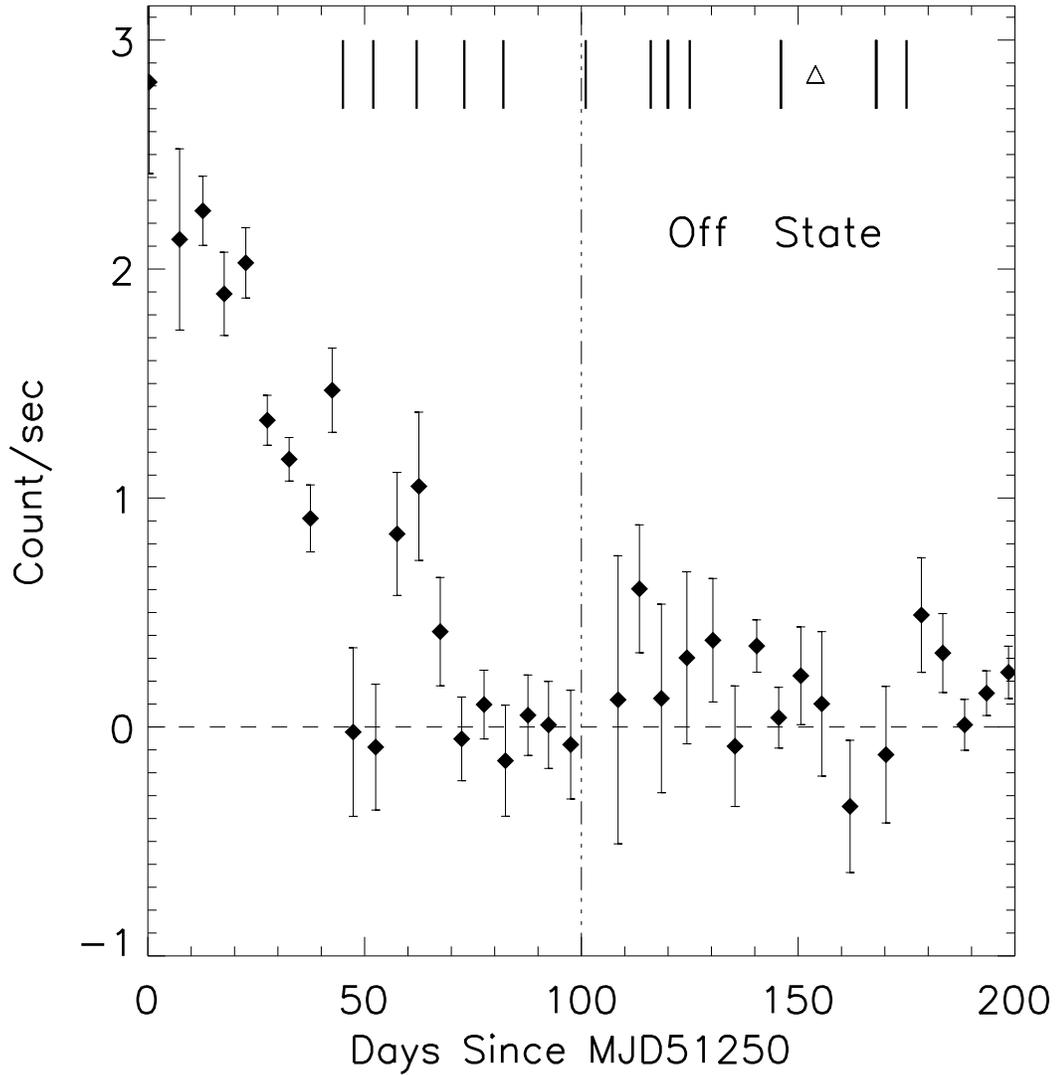}
\caption{Five-day averaged ASM light curve of GX~339$-$4. The lines at the 
top mark the times of the PCA observations and the triangle the time of 
the $BeppoSAX$ observation (see text). The dot-dashed line is drawn, 
somewhat arbitrarily, to indicate the beginning of the ``off'' state. 
MJD 51250 corresponds to 1999 Mar. 13. \label{fig1}}
\end{figure}

We observed GX~339$-$4 regularly with the Proportional Counter Array 
(PCA) and the High-Energy X-ray Time Experiment aboard RXTE between 
Apr. 27, 1999 and Sept. 4, 1999, as a part of our Target-of-Opportunity
campaign to follow the source from the ``low state'' to the ``off 
state''. For this investigation, we also included public data from other
observations of GX 339-4 that were made during this time period. A 
total of 17 observations were used to cover the very tail of the 
decaying phase of an X-ray outburst (which started in 1998). Fig.~1 
shows the ASM light curve of the source for this period and the times
of the RXTE observations. Since we were mostly interested in iron 
K$_{\alpha}$ line emission between 6--7 keV, we used data only from 
the PCA, which covers a nominal energy range of 2--60 keV. Table~1 
summarizes some of the key parameters of the observations.
\placefigure{fig1}
\placetable{table-1}

The PCA is consisted of five {\it Proportional Counter Units} (PCUs).
The operation of the detector requires that some of the PCUs be 
turned off at times, therefore, the number of PCUs in use varies
from observation to observation (as indicated in Table~1). To minimize 
instrumental effects, we chose to use data from the first (of the three)
xenon layer of each PCU, which is most accurately calibrated. Such
data also provides the best signal-to-noise ratio, because photons 
from real astronomical sources are mostly stopped in the first layer. 
This is clearly very important for studying iron K$_{\alpha}$ lines in 
BHCs, which are usually weak. The trade-off of such a choice is that 
we lost sensitivity at energies above $\sim$20 keV. This loss is not 
critical to this investigation, because the main objective here is, 
again, to study line emission at 6--7 keV and the energy band is 
broad enough that we can adequately constrain the underlying continuum.

We used ftools v5.0 to reduce and analyze the data, along with the most 
updated calibration files and background models that accompanied the 
release of the software package. Because the source became very faint
as it approached the ``off state'', to obtain adequate statistics we 
sometimes combined data from multiple observations (in which the X-ray
spectrum of the source did not vary significantly) to form one data 
group for further analyses. As a first step, we eliminated time 
intervals when the data is affected by earth occultation, SAA passages, 
and soft electron events. We then extracted a sum spectrum for each PCU
from the {\em Standard 2} data of individual or a group of observations, 
and constructed a corresponding background spectrum using the background 
models that are appropriate for the data set. For cases where the 
average source rate is greater than 40 ct/sec/PCU, a set of background 
models for bright sources (pca$_{-}$bkgd$_{-}$skyvle$_{-}$e4v20000131.mdl
and pca$_{-}$bkgd$_{-}$allskyactiv$_{-}$e4v20000131.mdl) were used; 
otherwise, those for faint sources 
(pca$_{-}$bkgd$_{-}$faintl7$_{-}$e4v19991214.mdl 
and pca$_{-}$bkgd${-}$faint240$_{-}$e4v19991214.mdl) were used. Finally,
we read both the sum and background spectra into XSPEC for visual
inspections. As a sanity check, we verified that the background spectrum
was in total agreement with the sum spectrum at the  highest energies
where no photons from the source are expected.

For spectral analysis, we constructed a response matrix for the first
xenon layer of each PCU (with pcarsp v2.43, which comes with ftools v5.0). 
To be conservative, we limited the analysis to an energy range of 3--20 
keV, in which the systematic uncertainty associated with the response 
matrices appears to be small and understood. Following the conventional 
approach, we added 1\% systematic error to the data to account for any 
remaining uncertainty in the calibration.
 
\section{Modeling and Results}
 
To model the observed spectrum of GX 339-4, we first experimented 
with a simple power law (with absorption). For each data set, we 
performed a joint fit in XSPEC to the spectra from all PCUs that were 
turned on during the observation(s). Note that in the cases of multiple 
observations combined the effective exposure time can be different for 
different PCUs, because some were turned off during a subset of 
observations (see Table 1). During the fit, we fixed the column density 
to $6 \times 10^{21}\mbox{ }cm^{-2}$ (Wilms et al. 1999; Kong  et al. 
2000), because it is poorly constrained by the data here. The model 
seems to adequately describe the spectrum in all cases, but we found 
significant residuals in the range of 6--7 keV (very similar to those 
shown in Fig.~2), indicating the presence of an iron K$_{\alpha}$ line. 
Adding a Gaussian 
function to the model indeed improves the fit significantly, for 
example, $\Delta \chi^2 = 90$ (with 3 added degrees of freedom) for 
the average spectrum of observations 6-17. However, the composite 
model (power law plus Gaussian) is still not satisfactory for early 
observations (when the source was relatively bright).

Next, we added a soft component, ``diskbb'' in XSPEC, to the 
``powerlaw+Gaussian'' model. This so-called ``multi-color disk'' 
model is commonly used to describe the low-energy ``excess'' of the 
observed X-ray continuum of BHCs. With the new model, we obtained
a significantly improved fit to the data for early observations. 
However, the improvement is minimal ($\Delta \chi^2$ = 2--4 with 
2 added degrees of freedom) for later observations. The results 
are summarized in Table~2. The uncertainties were derived by 
varying the parameter of interest until $\Delta \chi^2 = 2.7$ and 
thus represent 90$\%$ confidence regions (Lampton, Margon \& 
Bowyer 1976). 
\placetable{table-2}

The observed X-ray continuum of GX 339-4 is dominated by the power 
law component, which is typical of the ``low state'' of BHCs. It 
is worth pointing out that small inferred normalization for 
``diskbb'' must not be interpreted literally as small radius of 
the inner edge of the accretion disk, as is frequently done in the 
literature. In the context of Comptonization models, a weak disk 
component may simply mean that the X-ray emitting portion of the 
disk is covered up by the Comptonizing corona and thus little 
disk emission leaks through. The fact that the Comptonized 
emission (i.e., the power-law component) is dominant is consistent
with this scenario, if seed photons come mostly from the disk.

GX 339-4 shows a clear spectral evolution as it approaches the
``off state''. This is manifested in the gradual disappearance
of the soft component and the shift of the emission line. 
When the measured flux (between 3 and 20 keV) is greater than 
$\sim$5$\times 10^{-11}$ erg cm$^{-2} s^{-1}$, the best-fit
line energy is around 6.4 keV, which is in general agreement 
with the results of Wilms et al. (1999) based on earlier RXTE
observations of GX 339-4 in a similar spectral state. It should
be noted, however, that a direct comparison between our results
and those of Wilms et al. is difficult because of different
continuum models adopted. For these observations, we also fitted
the data with one of the models that Wilms et al. used (broken 
power law plus Gaussian) and the results are indeed similar to
theirs. In addition, we made an attempt to search for a possible 
absorption edge that may accompany the observed emission line, 
but we failed to detect it with any meaningful significance. In 
fact, the $\chi^2$ of the fit hardly changes with the addition 
of such an edge, regardless whether we fixed the energy of the 
edge or not. This may simply mean that the statistics of the data 
are not good enough to allow a detection of the edge. In general, 
we found that the energy of the line is {\it not} sensitive to 
the continuum models that we have tried, although both the width 
and flux of the line are.

When the source becomes fainter, however, the energy of the line 
moves up to around 6.7 keV, which has not been seen before. To 
quantify the significance of the line shift, we combined data 
from observations 6 to 17 (see Table~2) to obtain an average 
spectrum of the ``off state''. We then carried out the same
spectral analysis and the results are also shown in Table~2. We 
conducted the following F-test. We fixed the energy of the 
line to 6.4 keV in the model (with all other parameters floating) 
and fitted the data again. We found that $\chi^2$ increased by 15. 
The F-value is then 
$F=\frac{\Delta \chi^2/1}{\chi^2/dof}=\frac{15/1}{142/186} \simeq 20$, 
which corresponds to a null probability of $1.6\times 10^{-5}$. 
In other words, at a confidence level of 99.998\% the shift in
the line energy is real as the source entered the ``off state''.
For comparison, we also combined data from observations 1 to 5 to 
obtain an average spectrum of the ``low state'' and showed the 
results of spectral analysis in Table~2. The table also lists 
the equivalent width of the line for each data set. Again, 
our results roughly agree with those of Wilms et al. (1999) for 
cases in which the source is relatively bright. The line seems 
to strengthen toward the ``off state'', although the uncertainties 
are still large in these cases. 

To exclude the possibility that the observed line is a calibration 
artifact, we analyzed data from an observation of the Crab Nebula, 
which was made roughly during the same period (Observation ID 
40093-01-03-00, on Apr. 5, 1999). The data were screened, reduced, 
and analyzed by 
following exactly the same procedure as in the case of GX 339-4. 
We fitted the Crab spectrum with a simple power law (with absorption) 
and the best-fit parameters are: N$_{H}$=3.0$\times 10^{21}$ cm$^{-2}$, 
photon index 2.174, and 
normalization 12.14 $ph$ $keV^{-1}$ $cm^{-2}$ s$^{-1}$, all of which
appear normal for the source (e.g., Toor $\&$ Seward 1974). Fig.~2 
(upper panel) shows the residuals of the fit. Over the entire energy 
range of interest, {\em the residuals are never greater than 4\%}.
For comparison, we also plotted in Fig.~2 (lower panel) the residuals 
of a fit to the ``off state'' spectrum of GX 339-4 (see Table~2) 
without the Gaussian component. The peak of the line feature at 
$\sim$6.7 keV is about 40\% above the best-fit continuum model! 
Therefore, we conclude that the line is unlikely to be an artifact 
due to calibration uncertainties. 
\placefigure{fig2}

Next, we investigated whether the evolution of the emission line could
be caused by a systematic shift in the gain of the detector. We analyzed 
data from two observations of Cas~A: one was conducted during epoch 3 
(Observation ID 30804-01-01-00, on Apr. 14, 1998) and the other
(Observation ID 40806-01-04-00, on Sept. 15, 1999) was contemporary 
with the GX 339-4 observations. Again, we
followed the same procedure for data reduction and analysis. We found
an excellent agreement in the derived line energy between the two 
cases (and the results agree with those in the literature for Cas~A; 
e.g., Holt et al. 1994). We conclude that the observed evolution of 
the line energy is real.

Finally, we demonstrated that the detected line emission could not be due 
to contamination by diffuse X-ray emission in the ``Galactic Ridge''
(Koyama et al. 1986).  Integrated over the PCA's field of view, the 
flux of the iron K$_{\alpha}$ line from the Galactic Ridge emission
(at 6.7 keV) is about 1.7 -- 3.5 $\times 10^{-14}$ erg cm $^{-2}$s$^{-1}$ 
in the direction of GX~339$-$4 (Kaneda et al.1997; Valinia $\&$ Marshall 
1998). This is more than an order of magnitude smaller than that detected 
line flux of GX 339-4 
(5.0$^{+1.5}_{-1.2}$ $\times 10^{-13}$ erg cm $^{-2}$s$^{-1}$) in 
the `off' state. In other words, the contamination can be no more
than $\sim$10\%. 

\section{Discussion}
  
We have detected iron K$_{\alpha}$ line emission both in the ``low
state'' and the ``off state'' of GX 339-4. The line is centered at 
$\sim$6.4 keV when the source is relatively bright (in the ``low 
state'') but is shifted to $\sim$6.7 keV at lower fluxes (in the 
``off state''). GX 339-4 was also observed by BeppoSAX in the 
``off state'' (Kong et al. 2000; see Fig. 1 for the time of their 
observation), but no detection of any emission lines was reported (with 
an upper limit on the equivalent width of an iron K$_{\alpha}$ line 
about 600 eV). It should be noted, however, that compared to the PCA 
on RXTE (even with just the first xenon layer) the detectors on 
BeppoSAX (MECS in this case) have much smaller effective areas. 
Consequently, they (as well as the SIS or GIS on ASCA) are not well 
suited for studying relatively broad emission lines, which may be 
the case here.

For BHCs, iron K$_{\alpha}$ lines are usually attributed to the
fluorescent emission of iron species in the accretion disk which
is illuminated by a hard X-ray source. Therefore, the detection of 
such a line at $\sim$ 6.4 keV indicates that the accretion disk in 
GX 339-4 is neutral or only mildly ionized in the ``low state''.
As the source approaches the ``off state'', the X-ray flux 
decreases and thus the disk should be even less ionized (unless 
its density decreases drastically). Clearly, this line production 
mechanism would have difficulty in explaining the line detected 
at 6.7 keV in the ``off state''. Therefore, we speculate that 
the accretion flows in GX 339-4 must have undergone a significant 
change when the source went from the ``low state'' to the ``off 
state''. 

One possible scenario is that in the ``off state'', when the mass
accretion rate is presumably very low, the accretion may take place 
in the form of advection dominated accretion flows (ADAFs; e.g., 
Narayan \& Yi 1994). In an ADAF region, viscously dissipated energy 
during the accretion
process is mostly stored as internal energy of protons; very little 
is transfered to electrons, via inefficient Coloumb collisions, and 
is radiated away. Consequently, the temperature of the plasma can 
be very high ($10^{9-10}$ K) near the black hole but lower farther
away. Even in the outer region, gas is likely to be almost fully 
ionized. In this case, a recombination line can be produced at 6.7 
keV and 6.97 keV from He-like (Fe XXV) and H-like (Fe XXVI) iron 
ions, respectively, due to collisions (Masai 1984; 
Arnaud \& Rothenflug 1985). Calculations show that emission lines
from ADAFs are expected, although they should be very weak (Narayan
\& Raymond 1999). The line emission is enhanced in cases where an
outflow or wind is formed (known as advection-dominated inflow and 
outflow systems, or ADIOS; Blandford \& Begelman 1999). It is 
interesting to note that for GX 339-4 the observed strong correlation 
between the radio emission (and optical emission) and the hard X-ray 
emission has been suggested to be evidence for the presence of 
outflows in the ``low'' or ``off'' state (Fender et al. 1999). 

In our case, the measured equivalent width of the line, taken at 
face value, seems to be larger than expected (see
Narayan \& Raymond 1999), even for an ADIOS scenario. However, we 
emphasize, again, that the error bars are still very large and the 
values are sensitive to instrumental effects. We think that the very 
presence of the 6.7 keV line in the ``off state'' is 
interesting and seems to point to a physical origin like ADAF or 
ADIOS. If this can be proven, we would also have a physical 
distinction between the ``low state'' and the ``off state'': the 
transition between the two occurs when accretion flows switch from 
an optically thick, geometrically thin pattern to an optically thin, 
geometrically thick one (with possible outflows) and vice versa. 
In fact, the difference between the two states might have already
manifested itself observationally in the radio properties of 
GX 339-4: we noticed that the radio spectrum seems to be 
significantly steeper in the ``off state'' than in the ``low state''
(see Table 1 of Corbel et al. 2000). 

\acknowledgments
We thank J.H. You, Y.S. Yao, and X.L. Zhang for their carefully 
reading of an early version of the manuscript and for many useful 
discussions and suggestions. We also thank an anonymous referee 
for helpful comments.
We made use of the results provided by the ASM/RXTE teams and of the 
archival databases maintained by the High Energy Astrophysics Science 
Archive Research Center at NASA's Goddard Space Flight Center. This work 
was supported in part by NASA through LTSA grant NAG5-9998. S.N.Z and 
X.S. wish to acknowledge support from MSFC/NASA through contract NCC8-65 
and from GSFC/NASA through LTSA grants NAG5-7929 and NAG5-8523.

\clearpage
\begin{deluxetable}{lcccc}
\tablecolumns{5}
\tablewidth{0pc}
\tablecaption{PCA Observation Log. \label{table-1}}
\tablehead{
\colhead{No.}&\colhead{Obs. Id.}
&\colhead{Observation Date}&\colhead{PCUs}&\colhead{Exposure (s)\tablenotemark{\dag}}}
\startdata
1  & 40104--01--01--00 & 04/27/99  & 0,2,3 & 7680 \nl
2  & 40104--01--02--00 & 05/04/99  & 0,1,2,3,4 & 5792 \nl
3  & 40108--02--03--00 & 05/14/99  & 0,1,2,3,4 & 9856 \nl
4  & 40104--01--03--00 & 05/25/99  & 0,1,2,3 & 5152 \nl
5  & 40104--01--04--00 & 06/03/99  & 0,1,2,3,4 & 6128 \nl
6  & 40108--02--04--00 & 06/22/99  & 0,2,3,4 & 14608 \nl
7  & 40108--03--01--00 & 07/07/99  & 0,2,3 & 12896 \nl
8  & 40104--01--05--01 & 07/11/99  & 0,2,3,4 & 2256 \nl
9  & 40104--01--05--00 & 07/11/99  & 0,2,3,4 & 2704 \nl
10 & 40104--01--05--02 & 07/11/99  & 0,2,3,4 & 2432 \nl
11 & 40104--01--06--00 & 07/16/99  & 0,2,4 & 2752 \nl
12 & 40104--01--07--00 & 08/06/99  & 0,2 & 3072 \nl
13 & 40104--01--06--01 & 08/06/99  & 0,2,3 & 3040 \nl
14 & 40104--01--08--00 & 08/28/99  & 0,2,3 & 1056 \nl
15 & 40104--01--08--01 & 08/28/99  & 0,2,3 & 1984 \nl
16 & 40104--01--08--02 & 08/28/99  & 0,2,3 & 1456 \nl
17 & 40104--01--09--00 & 09/04/99  & 0,1,2,4 & 9024 \nl
\tablenotetext{\dag}{Total amount of on-source time after data filtering and screening}
\enddata
\end{deluxetable}

\clearpage
\begin{deluxetable}{lcccccccc}
\tablecolumns{9}
\tablewidth{0pc}
\tablecaption{Spectral properties of GX~339$-$4.\label{table-3}}
\tablehead{& \multicolumn{3}{c}{Gaussian}& \multicolumn{2}{c}{Multi-Color Disk } & \multicolumn{1}{c}{Power Law} & \\
\cline{2-4} \cline{5-6} \cline{7-7}\\
\colhead{No.} &\colhead{$E$} &\colhead{$\sigma_{K\alpha}$}&\colhead{$EW$}
&\colhead{T$_{in}$}&\colhead{N$_{d}$}&\colhead{$\alpha$}&\colhead{$\chi_{\nu}^{2}/dof$}&\colhead{Flux} \\
 & keV & keV & eV & keV & & & & }
\startdata
1     & $6.28^{+0.16}_{-0.30}$ & $0.86^{+0.33}_{-0.25}$ & $ 262^{+185}_{-90}$ 
      & $1.13^{+0.16}_{-0.18}$ & $ 2.5^{+2.6}_{-0.7}$ 
      & $1.43^{+0.03}_{-0.04}$ 
      & 131/116 & 4.25 \nl
2     & $6.46^{+0.09}_{-0.09}$ & $0.44^{+0.16}_{-0.18}$ & $133^{+41}_{-33}$ 
      & $1.40^{+0.12}_{-0.12}$ & $ 0.74^{+0.25}_{-0.11}$ 
      & $1.43^{+0.04}_{-0.05}$ 
      & 164/186 & 2.93 \nl
      
3     & $6.46^{+0.14}_{-0.21}$ & $0.53^{+0.31}_{-0.26}$ & $131^{+89}_{-45}$ 
      & $1.37^{+0.18}_{-0.19}$ & $ 0.32^{+0.30}_{-0.12}$ 
      & $1.56^{+0.07}_{-0.07}$ 
      & 140/186 & 1.52 \nl
      
4-5    & $6.47^{+0.15}_{-0.16}$ & $0.54^{+0.32}_{-0.28}$ & $297^{+189}_{-107}$ 
      & $1.13^{+0.20}_{-0.18}$ & $ 0.38^{+0.67}_{-0.20}$ 
      & $1.54^{+0.13}_{-0.16}$ 
      & 133/186 & 0.46 \nl
      
6-7  & $6.64^{+0.14}_{-0.19}$ & $ 0.3^{+0.6}_{-0.3}$ & $442^{+304}_{-149}$ 
      & $0.97^{+0.24}_{-0.23}$ & $ 0.36^{+0.63}_{-0.21}$ 
      & $1.65^{+0.18}_{-0.41}$ 
      & 111/147 & 0.15 \nl
      
8-13   & $6.83^{+0.16}_{-0.16}$ & $ 0.0^{+0.4}_{-0.0}$ & $595^{+248}_{-172}$ 
      & ... & ...
      & $2.17^{+0.16}_{-0.11}$ 
      & 86/149 & 0.07 \nl
      
14-17  & $6.67^{+0.13}_{-0.16}$ & $ 0.0^{+0.4}_{-0.0}$ & $502^{+199}_{-134}$ 
      & ... & ...
      & $2.16^{+0.06}_{-0.09}$ 
      & 106/188 & 0.11 \nl
\cline{1-9}
1-5  & $6.44^{+0.08}_{-0.07}$ & $ 0.53^{+0.13}_{-0.15}$ & $155^{+41}_{-34}$ 
        & $1.34^{+0.12}_{-0.09}$ & $ 0.60^{+0.20}_{-0.12}$ 
        & $1.46^{+0.02}_{-0.04}$ 
        & 196/186 & 1.91 \\
6-17  & $6.68^{+0.08}_{-0.11}$ & $ 0.0^{+0.4}_{-0.0}$ & $578^{+179}_{-142}$
        & $1.05^{+0.09}_{-0.23}$ & $ 0.27^{+0.30}_{-0.13}$
        & $1.51^{+0.50}_{-0.23}$ 
        & 142/186 & 0.13     

\enddata
\tablecomments{
The multi-color disk component is characterized by the temperature at
the inner edge of the disk ($T_{in}$) and the overall normalization
($N_{d}$, in units of ph cm$^{-2}$ s$^{-1}$). The energy,
width, and equivalent width of the emission line are denoted by
$E$, $\sigma_{K\alpha}$, and $EW$, respectively, and the power-law
photon index by $\alpha$. Also shown is the integrated 3--20 keV 
flux in units of 10$^{-10}$ ergs cm$^{-2}$s$^{-1}$. The error bars 
represent 90$\%$ confidence intervals. }
\end{deluxetable}
\clearpage
\clearpage
\begin{figure}
\plotone{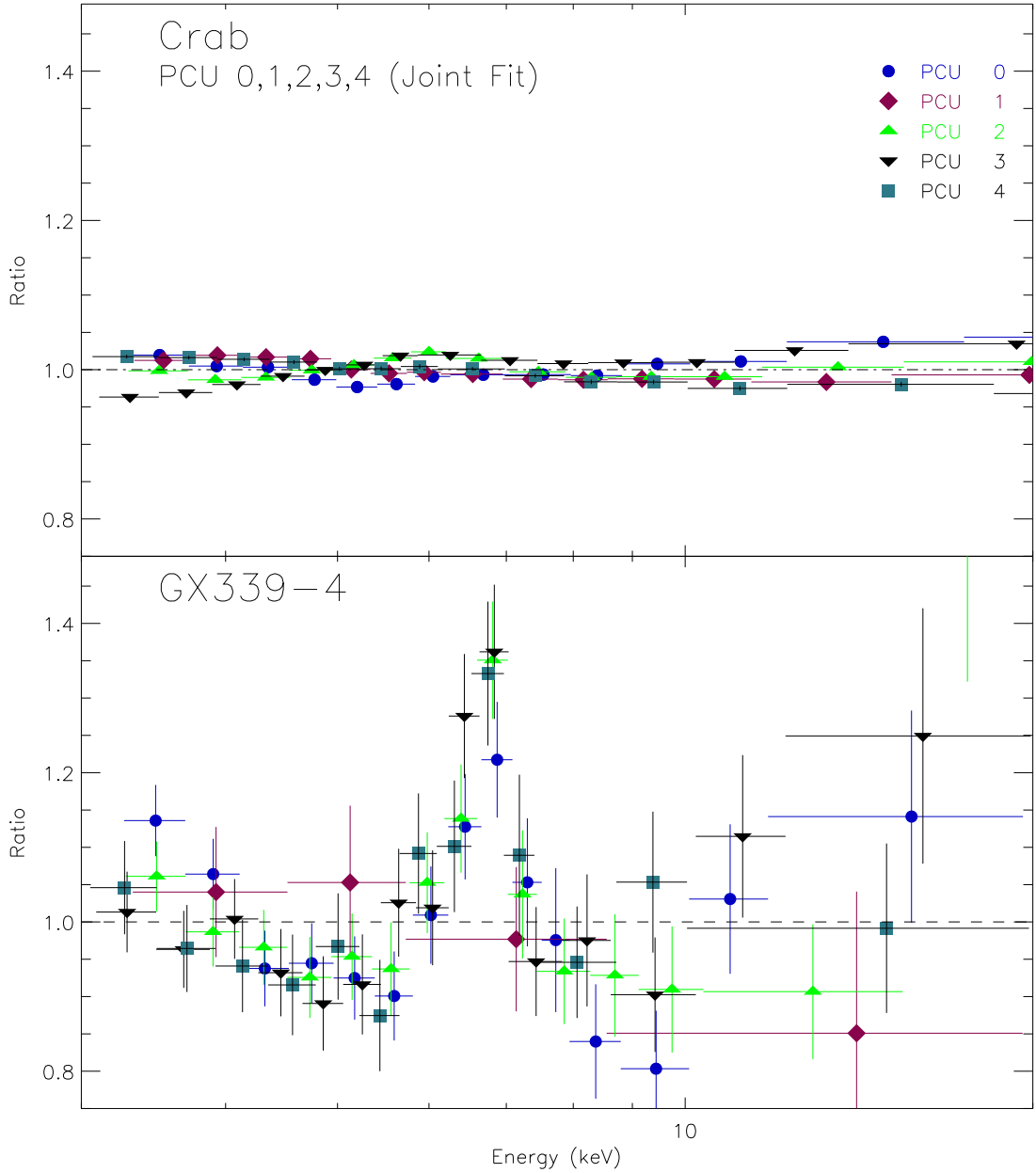}
\caption{Residual plots from the Crab and GX 339-4 observations. The 
plots of Crab are obtained by taking the ratio of the data to the best-fit 
power-law model. The plots of GX~339$-$4 are obtained by taking the ratio
of the data to the best-fit multi-color disk plus power-law model.
For GX~339$-$4, we combined the data from observations 6--17.
\label{fig3}}
\end{figure}

\end{document}